\documentclass[a4paper,twocolumn,english,pre,superscriptaddress,showpacs]{revtex4}
\usepackage[T1]{fontenc}
\usepackage[latin1]{inputenc}
\usepackage{graphicx}

\makeatletter

\def\vec#1{\mbox{\boldmath $#1$}}
\def\Bm{\textrm{Bm}}

\usepackage{babel}
\makeatother
\begin{document}

\title{Edge pinch instability of liquid metal sheet in a transverse high-frequency
ac magnetic field}

\author{J\={a}nis Priede}

\email{priede@sal.lv}

\affiliation{Institute of Physics, University of Latvia, LV--2169 Salaspils, Latvia}

\author{Jacqueline Etay}

\author{Yves Fautrelle}

\affiliation{CNRS INPG EPM-Madylam, ENSHMG BP 95 St Martin d'Hères cedex, France}

\begin{abstract}
We analyze the linear stability of the edge of a thin liquid metal
layer subject to a transverse high-frequency ac magnetic field. The
layer is treated as a perfectly conducting liquid sheet that allows
us to solve the problem analytically for both a semi-infinite geometry
with a straight edge and a thin disk of finite radius. It is shown
that the long-wave perturbations of a straight edge are monotonically
unstable when the wave number exceeds the critical value $k_{c}=F_{0}/(\gamma l_{0}),$
which is determined by the linear density of the electromagnetic force
$F_{0}$ acting on the edge, the surface tension $\gamma,$ and the
effective arclength of edge thickness $l_{0}.$ Perturbations with
wavelength shorter than critical are stabilized by the surface tension,
whereas the growth rate of long-wave perturbations reduces as $\sim k$
for $k\rightarrow0.$ Thus, there is the fastest growing perturbation
with the wave number $k_{\max}=2/3k_{c}.$ When the layer is arranged
vertically, long-wave perturbations are stabilized by the gravity,
and the critical perturbation is characterized by the capillary wave
number $k_{c}=\sqrt{g\rho/\gamma}$, where $g$ is the acceleration
due to gravity and $\rho$ is the density of metal. In this case,
the critical linear density of electromagnetic force is $F_{0,c}=2k_{c}l_{0}\gamma,$
which corresponds to the critical current amplitude $I_{0,c}=4\sqrt{\pi k_{c}l_{0}L\gamma/\mu_{0}}$
when the magnetic field is generated by a straight wire at the distance
$L$ directly above the edge. By applying the general approach developed
for the semi-infinite sheet, we find that a circular disk of radius
$R_{0}$ placed in a transverse uniform high-frequency ac magnetic
field with the induction amplitude $B_{0}$ becomes linearly unstable
with respect to exponentially growing perturbation with the azimuthal
wave number $m=2$ when the magnetic Bond number exceeds $\Bm_{c}=B_{0}^{2}R_{0}^{2}/(2\mu_{0}l_{0}\gamma)=3\pi.$
For $\Bm>\Bm_{c},$ the wave number of the fastest growing perturbation
is $m_{\max}=\left[2\Bm/(3\pi)\right].$ These theoretical results
agree well with the experimental observations. 
\end{abstract}


\maketitle

\section{Introduction}

In several induction melting processes, such as the cold crucible
or electromagnetic levitation, liquid metal with a free surface is
subject to ac magnetic fields that may cause considerable deformations
of liquid metal resulting from the electromagnetic forces due to the
eddy currents, which are often confined in a thin skin layer beneath
the surface \cite{Sneyd-1993}. It has been observed that the free
surface sometimes may become strongly asymmetric and even irregular
when a sufficiently strong magnetic field is applied \cite{Perrier-etal-2003,Fautrelle-etal-2005,Mohring-etal-2005}. 

Most of theoretical studies of the effect of ac magnetic field on
the stability liquid metal surfaces have been concerned with flat
surfaces subject to tangential uniform magnetic field. McHale and
Melcher \cite{McHale-Melcher-1982} were the first to show that the
time-averaged electromagnetic force has a destabilizing effect giving
rise to traveling waves on the surface of liquid metal. Although the
theoretical instability threshold is in good agreement with experimental
results, the predicted growth rates are too small compared to the
experimental observations. Note that such small growth rates are typical
for the electromagnetic instabilities caused by the currents induced
by motion of conducting media in ac magnetic fields \cite{Priede-Gerbeth-2005}.
This simple model was revisited by a number of authors using various
approximations. First, Garnier and Moreau \cite{Garnier-Moreau-1983}
found that ac magnetic field has only a stabilizing effect on the
surface waves when the currents induced by metal flow are neglected.
Deepak and Evans \cite{Deepak-Evans-1995} took into account the motion
of a surface but not the associated flow in the liquid, although both
have a comparable effect, and they concluded that ac magnetic field
can, however, give rise to unstable traveling surface waves. Stability
of a flat metal layer suspended by means of a uniform magnetic field,
as in the experiment of Hull \textit{et al.} \cite{Hull-etal-1989},
has been studied by Ramos and Castellanos \cite{Ramos-Castellanos-1996},
who analyzed the effect of the viscosity on Rayleigh-Taylor type instability,
which is unavoidable in this system. Fautrelle and Sneyd \cite{Fautrelle-Sneyd-1998}
used a more elaborate model, taking into account not only the time-averaged
but also the oscillating part of the electromagnetic force, which
results in much stronger parametric instabilities when the frequency
of surface waves is sufficiently close to the multiple of the electromagnetic
force frequency. Note that this simple model of a flat surface with
tangential magnetic field leads only to traveling but not stationary
wave instabilities, which require consideration of nonplanar surfaces
in nonuniform magnetic fields. A stability analysis was performed
by Karcher and Mohring \cite{Karcher-Mohring-2003} to describe the
experimental observations of static surface instabilities by Mohring
\textit{et al.} \cite{Mohring-etal-2005}. However, drastic simplifications
were made in the latter analysis. First, the authors used the mirror
image method to find the magnetic field distribution at the end of
annular gap filled with liquid metal, however this method is applicable
only to simple geometries, such as half-space or a sphere \cite{Jackson-75}.
Second, they neglect the effect of surface perturbation on the magnetic
field distribution, although the coupling between both constitutes
the basic mechanism of this instability.

\begin{figure}
\begin{center}\includegraphics[%
  width=0.75\columnwidth]{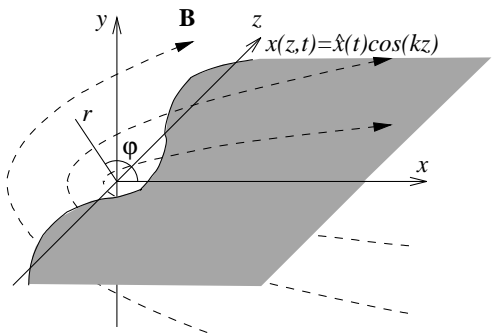}\end{center}

\caption{\label{cap:sketch}Sketch to the formulation of the problem with
the magnetic flux lines (dashed) bending around perturbed edge $x(z,t)=\hat{x}(t)\cos(kz).$}
\end{figure}

In this work, we propose a simple theoretical model to describe such
static surface instabilities. The model consists of a flat liquid
metal layer in a transverse ac magnetic field. The ac frequency is
assumed to be high so that the magnetic field is effectively expelled
from the layer by the skin effect. The layer is assumed to be thin
so that it can be regarded as a liquid perfectly conducting sheet.
We start with the linear stability analysis of the straight edge of
a semi-infinite liquid sheet, which allows us to work out a general
approach, which is applied later to the liquid layer in the magnetic
field of a straight wire parallel to the edge and to a thin circular
liquid disk in a uniform transverse magnetic field. We describe a
pinch-type instability of the edge of a liquid metal sheet with the
following mechanism. The magnetic field bends around the edge of a
perfectly conducting sheet, giving rise to the magnetic pressure on
the edge, which tries to compress the sheet laterally. In the case
of a straight edge, the uniform magnetic pressure along the edge is
balanced by a constant pressure in the sheet. A wavelike perturbation
of the edge causes the magnetic flux lines to diverge at wave crests
and convergence at troughs. This redistribution is because the magnetic
flux lines along the sheet are perpendicular to the electric current
lines that are directed along the edge and, thus, the magnetic flux
lines have to be perpendicular to the latter. As the result, the magnetic
pressure is reduced at the crests and increased at the troughs, which
enhances the perturbation. 

The paper is organized as follows. In Sec. II, we consider the general
model of a semi-infinite perfectly conducting liquid sheet with a
straight edge, and compare with the experiment. In Sec. III, the linear
stability problem for a thin disk in a transverse magnetic field is
solved and compared with experiment. The results are summarized in
Sec. IV.

\section{perfectly conducting liquid sheet}

\subsection{Mathematical model}

Consider a thin layer of liquid metal submitted to a transverse ac
magnetic field $\vec{B}$. We assume that the layer is semi-infinite
and lies on the right-hand side of $x$-$z$ plane, so that the unperturbed
edge of the layer coincides with $z$ axis, as shown in Fig. \ref{cap:sketch}.
Field frequency is assumed to be high, so that the layer is effectively
impermeable to the magnetic field because of the skin effect. In addition,
the layer is assumed to be a thin sheet in a static equilibrium state
supported by a flat horizontal non-wetting surface or constrained
in the gap between two parallel walls. Note that this model represents
a special case of a thin superconductor film (see, e.g., \cite{Brandt-1994}).
The magnetic field $\vec{B}$ in the free space around the sheet can
be described by the scalar magnetic potential $\Psi.$ Then the solenoidity
constraint $\vec{\nabla}\cdot\vec{B}=0$ results in \begin{equation}
\vec{\nabla}^{2}\Psi=0.\label{eq:Lapl-Psi}\end{equation}
The impermeability condition at both sides of the sheet is \begin{equation}
\left.\left(\vec{n}\cdot\vec{B}\right)\right|_{y=\pm0;\, x>0}=\left.\frac{\partial\Psi}{\partial n}\right|_{y=\pm0;\, x>0}=0,\label{eq:bnd-cnd}\end{equation}
 where $\vec{n}$ is the surface normal vector. First, we focus on
the distribution of the magnetic field in the vicinity of the edge,
which can conveniently be described in the cylindrical coordinates
with the $z$ axis coinciding with the edge and the polar angle $\varphi$
measured from the $x$ axis, as illustrated in Fig. \ref{cap:sketch}.
The solution for the unperturbed potential in the vicinity of straight
edge satisfying condition (\ref{eq:bnd-cnd}) is \cite{Landau-1984}
\begin{equation}
\Psi_{0}(r,\varphi)=C_{0}\sqrt{r}\cos(\varphi/2),\label{eq:Psi0}\end{equation}
 where $C_{0}$ is an unknown constant. According to simple dimensional
arguments, determining $C_{0}$ requires an external length scale
that, however, is missing in this simple model. Thus, $C_{0}$ can
be determined for a strip of finite width, a magnetic field generated
by a straight wire placed at some distance parallel to the edge or
a circular disk of finite size, that will be done in the following
section. But first, we develop a general approach without specifying
$C_{0}.$ 

Suppose that there is a perturbation of the edge position $x=x_{1}(z,t)=\hat{x}(t)\cos(kz)$
with a small, generally time-dependent amplitude $\hat{x}(t)$ and
the wave number $k$ along the $z$ axis. This perturbation gives
rise to the potential perturbation that can be presented as\[
\Psi(r,\varphi,z)=\Psi_{0}(r,\varphi)+\varepsilon\Psi_{1}(r,\varphi,z)+\cdots,\]
where $\Psi_{1}$ is a perturbation with small amplitude $\varepsilon.$
To relate the perturbation of a potential to that of the edge, we
need an additional condition at the edge, which is derived as follows.
For the surface current with density $\vec{J}$, we have $\mu_{0}\vec{J}=\vec{n}\times\vec{B}$
\cite{Landau-1984}, where $\mu_{0}$ is the permeability of vacuum.
According to this relation, the magnetic field along the sheet is
perpendicular to the current. Consequently, the magnetic field has
be to perpendicular to the edge because the current is flowing along
the latter. Thus along the edge $L,$ we obtain $\left.\left(\vec{\tau}\cdot\vec{B}\right)\right|_{L}=\left.\partial\Psi/\partial\tau\right|_{L}=0,$
where $\vec{\tau}$ is the unit vector tangential to the edge. This
condition in turn implies that$\left.\Psi\right|_{L}=\mathrm{\mathit{const}},$
where $\mathrm{\mathit{const}}=0$ may be chosen because the potential
is defined up to an additive constant. Applying this condition at
the perturbed edge, we obtain up to the first-order terms in the perturbation
amplitude \[
\left.\Psi\right|_{x=x_{1}}\approx\left.\left(\Psi_{0}+\frac{\partial\Psi_{0}}{\partial x}x_{1}+\varepsilon\Psi_{1}\right)\right|_{r\rightarrow0;\,\varphi=0}=0,\]
 which results in \begin{equation}
\left.\varepsilon\Psi_{1}\right|_{r\rightarrow0;\,\varphi=0}=-\left.\frac{\partial\Psi_{0}}{\partial x}x_{1}\right|_{r\rightarrow0;\,\varphi=0}=-\left.\frac{C_{0}}{2}\frac{x_{1}}{\sqrt{r}}\right|_{r\rightarrow0}.\label{eq:cnd-edg}\end{equation}
Note that the base potential cannot formally be expanded in power
series directly at the edge because the necessary derivative is singular
there. To circumvent this, we take the expansion not exactly at the
edge but define it as a limit when the expansion point approaches
the edge.

\begin{figure}
\begin{center}\includegraphics[%
  bb=125bp 95bp 355bp 250bp,
  width=1\columnwidth]{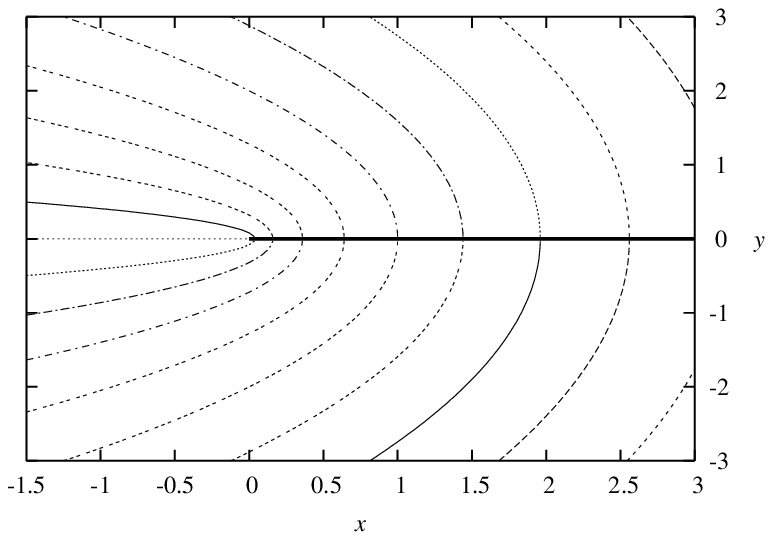}\put(-240,0){(a)}\end{center}

\begin{center}\includegraphics[%
  bb=125bp 95bp 355bp 250bp,
  width=1\columnwidth]{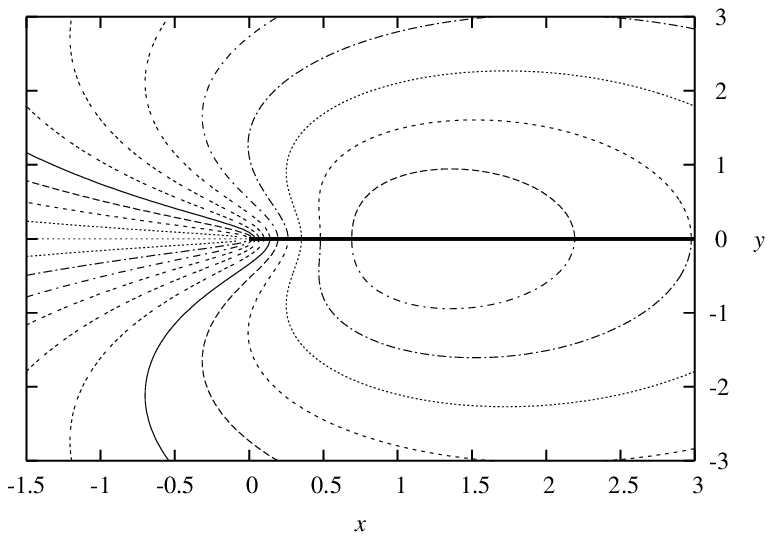}\put(-240,0){(b)}\end{center}

\caption{\label{cap:Psi}Scalar magnetic potential in the vicinity of a straight
edge (a) and its perturbation amplitude with $1/k$ used as the length
scale (b). }
\end{figure}

\begin{figure}
\begin{center}\includegraphics[%
  width=0.5\columnwidth]{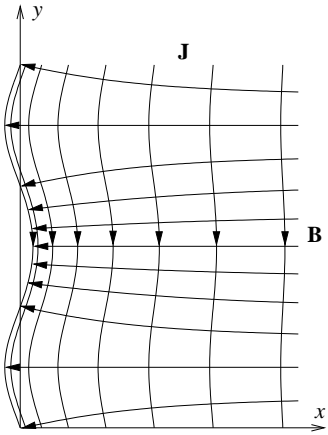} \end{center}

\caption{\label{cap:layer}Magnetic flux and current lines along the layer
in the vicinity of the perturbed edge.}
\end{figure}

The potential perturbation that satisfies Eq. (\ref{eq:bnd-cnd})
is of the same form as the base field $\hat{\Psi}_{1}(r,\varphi)=f_{1}(r)\cos(\varphi/2).$
When substituted into Eq. (\ref{eq:Lapl-Psi}), this leads to \[
\frac{1}{r}\frac{d}{dr}\left(r\frac{df_{1}}{dr}\right)+\frac{1}{4}\frac{f_{1}}{r^{2}}-k²f_{1}=0.\]
The solution satisfying Eq. (\ref{eq:cnd-edg}) is $f_{1}(r)=-\frac{1}{2}C_{0}x_{1}e^{-kr}/\sqrt{r},$
and the full potential including the base field is \begin{equation}
\Psi(r,\varphi,z)=C_{0}\sqrt{r}\left(1-\frac{1}{2}\frac{x_{1}}{r}e^{-kr}\cos(kz)\right)\cos(\varphi/2).\label{eq:sol-Psi}\end{equation}
Note that the potential above, which is defined relative to the unperturbed
edge, contains a singularity at the unperturbed edge. This singularity
can be removed by proceeding to the coordinates defined relative to
the perturbed edge as $\vec{r}=\vec{r}'+\vec{e}_{x}x_{1},$ and expanding
the solution in terms of $x_{1}.$ Thus, we obtain up to the first-order
terms in the perturbation amplitude \begin{eqnarray}
\Psi(\vec{r}'+\vec{e}_{x}x_{1}) & \approx & \left.\left(\Psi_{0}+\frac{\partial\Psi_{0}}{\partial x}x_{1}+\varepsilon\Psi_{1}\right)\right|_{\vec{r}=\vec{r}'}\label{eq:Psi'}\\
 & = & C_{0}\sqrt{r'}\left[1+\frac{x_{1}}{2r'}\left[1-e^{-kr'}\cos(kz)\right]\right],\nonumber \end{eqnarray}
where $r'$ is the cylindrical radius relative to the perturbed edge.
Having no singularity anymore, this solution simplifies in the vicinity
of the edge to \[
\left.\Psi\right|_{r'\rightarrow0}\approx\tilde{C}_{0}\sqrt{r'}\cos(\varphi/2),\]
 where $\tilde{C}_{0}=C_{0}\left[1+\frac{1}{2}\hat{x}k\cos(kz)\right].$
Thus perturbation of the edge results just in a redefinition of the
constant $C_{0},$ which is now replaced by $\tilde{C}_{0},$ whereas
the distribution of the potential remains the same as for the straight
edge obtained above. This is because a smoothly perturbed edge looks
straight again when examined on a sufficiently small scale. The scalar
magnetic potential in the vicinity of a straight edge and its perturbation
amplitude defined by Eq. (\ref{eq:Psi'}) are plotted in Fig. \ref{cap:Psi}
with $1/k$ used as the length scale. The corresponding magnetic flux
and current lines along the layer in the vicinity of the perturbed
edge are shown in Fig. \ref{cap:layer}. As seen, the magnetic flux
lines diverge at wave crests and converge at troughs in order to remain
perpendicular to the edge, as discussed above. This redistribution
of the magnetic flux lines at the perturbed edge is the principal
physical mechanism behind the instability considered in this study.


\begin{figure}
\begin{center}\includegraphics[%
  width=0.5\columnwidth]{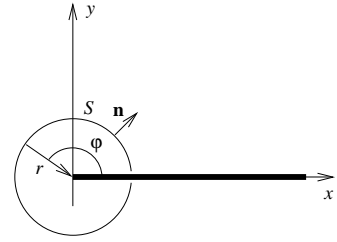}\end{center}

\caption{\label{cap:edge} Evaluation of the electromagnetic force on the
edge by integration of the Maxwell stress tensor over a small cylindrical
surface $S$ enclosing the edge.}
\end{figure}

In the perfect conductor approximation, the electromagnetic force
due to an ac magnetic field reduces to an effective magnetic pressure
acting on the surface of the layer with the time-averaged value $p_{m}=\left|\vec{B}_{0}\right|^{2}/(4\mu_{0}),$
where $\vec{B}_{0}$ is the amplitude of an ac magnetic field. Note
that part of the magnetic pressure, which oscillates with double ac
frequency, is neglected here by assuming the frequency to be so high
that inertia precludes any considerable reaction of the liquid. According
to Eq. (\ref{eq:sol-Psi}), the magnetic pressure increases toward
the edge as $\sim1/r$ and, thus, it becomes singular at $r=0$. Nevertheless,
the integral force on the edge has a finite value. This is because
the magnetic pressure at the edge of a layer of small but finite thickness
$\sim d_{0}$ increases as $\sim1/d_{0},$ which, integrated over
the thickness, results in a finite value independent of $d_{0}$.
The force on the edge can be evaluated by integrating the Maxwell
stress tensor over a small cylindrical surface $S$ enclosing the
edge, as shown in Fig. \ref{cap:edge}, that results in the first-order
terms in\begin{eqnarray}
\vec{F} & = & \frac{1}{2\mu_{0}}\lim_{r\rightarrow0}\int_{0}^{2\pi}\left[-\frac{1}{2}\vec{B}^{2}\vec{n}+(\vec{B}\cdot\vec{n})\vec{B}\right]rd\varphi\nonumber \\
 & = & \vec{e}_{x}(F_{0}+F_{1}),\label{eq:frc}\end{eqnarray}
 where \begin{equation}
F_{0}=\frac{\pi C_{0}^{2}}{8\mu_{0}}\label{eq:F0}\end{equation}
 and $F_{1}=F_{0}k\hat{x}\cos(kz)$ are the base force and its perturbation,
respectively.

Further, we assume the sheet to be an inviscid liquid, and consider
a small-amplitude potential flow associated with the edge perturbation.
Justification of this assumption, particularly at the threshold of
monotonous instability, is discussed at the end of the section. Then
the linearized Euler equation applied to a potential velocity field
$\vec{v}=\vec{\nabla}\Phi,$ \[
\rho\frac{\partial\vec{v}}{\partial t}+\vec{\nabla}p=\vec{\nabla}\left(\rho\frac{\partial\Phi}{\partial t}+p\right)=0,\]
leads to the pressure distribution in the sheet $p=p_{0}-\rho\partial\Phi/\partial t=p_{0}+p_{1},$
where $p_{0}$ is a constant base pressure and $p_{1}=-\rho\partial\Phi/\partial t$
is the perturbation of pressure. Velocity potential $\Phi$ is governed
by the incompressibility constraint $\vec{\nabla}\cdot\vec{v}=0$
resulting in\begin{equation}
\vec{\nabla}^{2}\Phi=0.\label{eq:Phi}\end{equation}
 We integrate the normal stress balance over the edge by assuming
both the pressure and curvature to be constant because of the small
thickness of the layer, which yields \begin{equation}
\left.p\right|_{x=0}=\frac{\gamma}{R}+\frac{F}{l_{0}},\label{eq:pressure}\end{equation}
where $l_{0}$ is the effective arclength of the edge thickness; $\gamma$
is the surface tension and $1/R$ denotes the curvature of the edge.
For an unperturbed edge, we have $p_{0}=\gamma/R_{0}+F_{0}/l_{0}.$
Then for the perturbation, the balance condition takes the form\begin{equation}
-\rho\left.\frac{\partial\Phi}{\partial t}\right|_{x=0}=\frac{\gamma}{R_{1}}+\frac{F_{1}}{l_{0}},\label{eq:nrm-blnc}\end{equation}
where $1/R_{1}\approx\vec{\nabla}^{2}x_{1}$ is the curvature perturbation
of the edge. In addition, we have a kinematic constraint\begin{equation}
\left.v_{x}\right|_{x=0}=\left.\frac{\partial\Phi}{\partial x}\right|_{x=0}=\frac{\partial x_{1}}{\partial t}.\label{eq:kinem}\end{equation}
\begin{figure}
\begin{center}\includegraphics[%
  width=1\columnwidth]{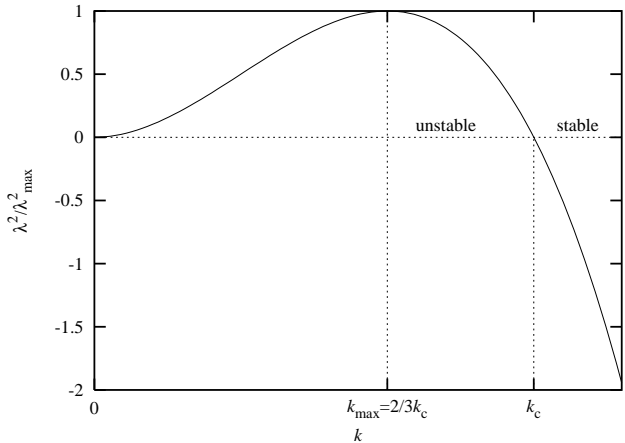}\end{center}

\caption{\label{cap:grate}Characteristic dependence of growth rate square
scaled with its maximum value on the wave number.}
\end{figure}
Now we search for the amplitude of edge perturbation of the form $\hat{x}(t)=x_{0}e^{\lambda t},$
where $\lambda$ is, in general, a complex growth rate, whose real
part has to be negative for the perturbation to be stable. The hydrodynamic
potential is of the form \[
\Phi(x,z,t)=\hat{\Phi}(x)\cos(kz)e^{\lambda t},\]
which when substituted into Eq. (\ref{eq:Phi}) leads to $d^{2}\hat{\Phi}/dz^{2}-k^{2}\hat{\Phi}=0,$
whose solution decaying away from the edge is $\hat{\Phi}(x)=\Phi_{0}e^{-kx}.$
The amplitude of the hydrodynamic potential is related to that of
the edge perturbation by the kinematic constraint (\ref{eq:kinem})
$\Phi_{0}=-x_{0}\lambda/k.$ Finally, the normal stress balance (\ref{eq:nrm-blnc})
yields the growth rate depending on the wave number\begin{equation}
\lambda(k)=k\sqrt{\frac{1}{\rho}\left(\frac{F_{0}}{l_{0}}-k\gamma\right)},\label{eq:grate-edge}\end{equation}
which implies that long-wave perturbations with the wave numbers $0<k<k_{c}=F_{0}/(\gamma l_{0})$
have positive growth rates and, thus, they are unstable, as illustrated
in Fig. \ref{cap:grate}. The stronger the electromagnetic force $F_{0}$
on the edge, the shorter the critical wavelength. The waves that are
shorter than the critical one are stabilized by the surface tension.
Although long waves are always unstable, their growth rate reduces
as $\sim k$ for $k\rightarrow0.$ Thus there is a perturbation with
$k_{\max}=\frac{2}{3}k_{c}$ for which the growth rate attains the
maximum $\lambda_{\max}=k_{\max}\sqrt{F_{0}/(3\rho l_{0})}$ (see
Fig. \ref{cap:grate}). 

Concerning the effect of neglected viscosity, simple physical arguments
suggest that this instability can only be slowed down but not prevented
by the viscosity. Note that the viscosity is inherently related to
the fluid flow. But at the threshold of monotonous instability, where
$\lambda(k_{c})=0,$ the characteristic time of monotonous instability
tends to infinity. Consequently, there is no characteristic time and,
thus, no characteristic velocity scale for the monotonous marginally
stable mode which, therefore, cannot be affected by the viscosity.

\subsection{Comparison with experiment}

\begin{figure}
\begin{center}\includegraphics[%
  bb=125bp 110bp 350bp 350bp,
  width=0.9\columnwidth]{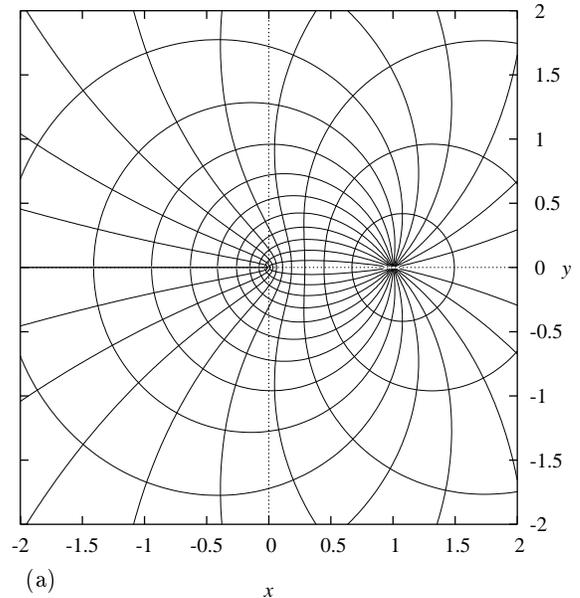}\put(-210,10){(a)}\end{center}

\begin{center}\includegraphics[%
  bb=125bp 110bp 350bp 350bp,
  width=0.9\columnwidth]{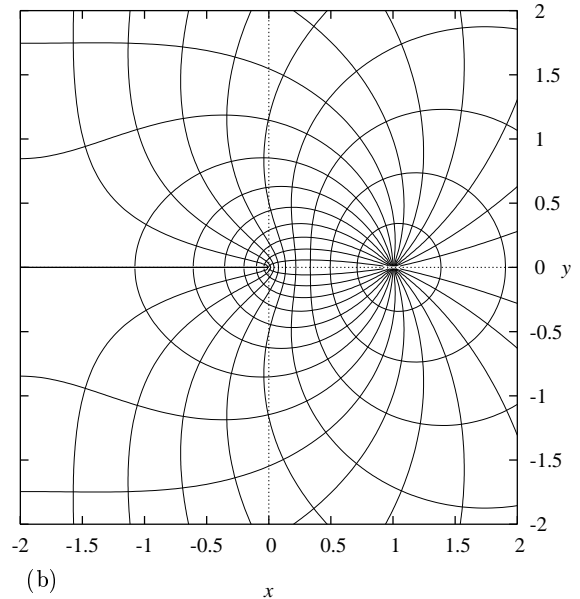}\put(-210,10){(b)}\end{center}

\caption{\label{cap:edge-wire}Magnetic flux lines and isolines of the scalar
magnetic potential represented by the real and imaginary parts of
the complex potential for a straight wire placed at $x=L=1$ and a
semi-infinite sheet at $x<0$ (a) and for the sheet of finite height
$-H<x<0$ transverse to a perfectly conducting wall at $x=-H=-2$
(b).}
\end{figure}

To compare our theory with the experiment of Mohring \textit{et al.}
\cite{Mohring-etal-2005}, similarly to \cite{Karcher-Mohring-2003},
we unfold the annular layer of InGaSn (Galinstan) melt used in experiment
by approximating it by a semi-infinite flat perfectly conducting liquid
sheet. Magnetic field is approximated by that of a straight wire lying
at the distance $L$ from the edge in the plane of the sheet. The
distance $L$ provides us with the length scale necessary to specify
the constant $C_{0}$ used in our model above. This constant follows
from the complex potential of the magnetic field, which is obtained
by the conformal mapping \[
w(\zeta)=\frac{\mu_{0}I_{0}}{2\pi}\log\frac{\sqrt{\zeta}+\sqrt{L}}{\sqrt{\zeta}-\sqrt{L}},\]
where $\zeta=x+iy,$ as $C_{0}=\mu_{0}I_{0}/(\pi\sqrt{L}).$ The magnetic
flux lines and isolines of the scalar magnetic potential represented
by the real and imaginary part of this complex potential are shown
in Fig. \ref{cap:edge-wire}(a) with $L$ used as the length scale.
In addition, we consider the gravity with the acceleration $\vec{g}=\vec{e}_{x}g$
directed downwards along the sheet normally to its edge. Then Eqs.
(\ref{eq:grate-edge}) and (\ref{eq:F0}) with $C_{0}$ defined above
result in \begin{equation}
\lambda(k)=k\sqrt{\frac{1}{\rho}\left(\frac{\mu_{0}I_{0}^{2}}{8\pi L_{0}l_{0}}-k\gamma\right)-\frac{g}{k}}.\label{eq:disp-rel}\end{equation}

\begin{figure}
\begin{center}\includegraphics[%
  width=1\columnwidth]{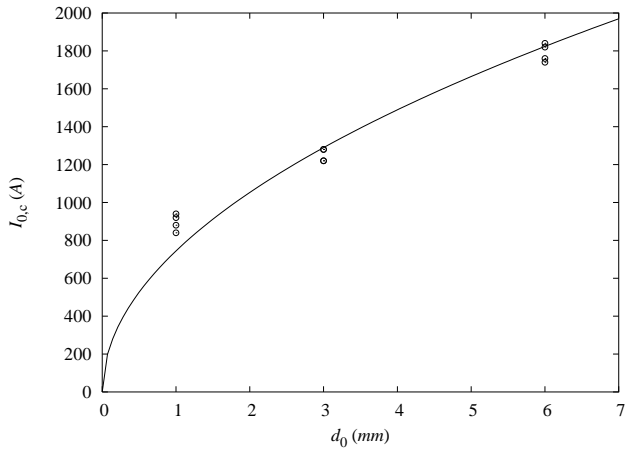}\end{center}

\caption{\label{cap:exp-edge}Critical current amplitude vs the layer thickness.
The circles correspond to the measured total current amplitude for
the gap $L_{\mathrm{exp}}=17\,\mathrm{mm}$ between the metal surface
and the lower side of the inductor \cite{Mohring-etal-2005}. The
theoretical curve corresponds to the best fit with the effective gap
with $L\approx45\,\mathrm{mm}$ for a semi-infinite sheet and $L\approx28\,\mathrm{mm}$
for a sheet of finite height $H=24\,\mathrm{mm}$ with a perfectly
conducting bottom. }
\end{figure}

As is easy to see, the gravity stabilizes long-wave disturbances,
whereas short waves are stabilized by the surface tension. Thus, the
first unstable mode, defined by $\lambda(k_{c})=0,$ appears at the
capillary wave number $k_{c}=\sqrt{\frac{g\rho}{\gamma}}\approx0.296\,\mathrm{mm}^{-1}$
that corresponds to the critical wavelength $\Lambda_{c}=2\pi/k_{c}\approx21.2\,\mathrm{mm},$
where $\rho=6440\,\mathrm{kq}/\mathrm{m}^{3}$ and $\gamma=0.718\,\mathrm{N}/\mathrm{m}$
are the density and surface tension of Galinstan \cite{Mohring-etal-2005}.
Note that in the experiment, the surface of liquid metal is covered
by a layer of NaOH solution. Thus, perturbation of the hydrostatic
pressure at the interface is determined by the density difference
of GaInSn and NaOH. Assuming the latter to have the density of water,
we find the critical wavelength $\Lambda_{c}\approx23\,\mathrm{mm}$
that coincides very well with that of the static surface deformation
observed in the experiment. The critical electromagnetic force follows
from Eq. (\ref{eq:disp-rel}) as $F_{0,c}=\sqrt{2\gamma l_{0}}/k_{c},$
which corresponds to the critical current amplitude \begin{equation}
I_{0,c}=\pi\sqrt{\frac{8d_{0}L\sqrt{g\rho\gamma}}{\mu_{0}}},\label{eq:I0-c}\end{equation}
where the edge arclength $l_{0}$ over the layer thickness $d_{0}$
is approximated by a half-circle, i.e., $l_{0}=\pi d_{0}/2.$ In order
to compare this result with the measured critical currents\cite{Mohring-etal-2005},
note that the coil used in the experiment consists of two horizontal
layers, each of which contains five windings. Thus, the measured current
has to be multiplied by 10 to obtain the total current amplitude.
Unfortunately, the authors do not specify the coil dimensions but
give only the gap width $L_{\mathrm{exp}}=17\,\mathrm{mm}$ between
the metal surface and the lower side of the inductor which is not
sufficient for comparison with our theory. Therefore we treat the
distance $L$ as a free parameter to fit the experimental results
that yield $L\approx45\,\mathrm{mm}$ (see Fig. \ref{cap:exp-edge}).
Note that the model of a semi-infinite layer may not be very adequate
for the given experiment with the layer extention $H=24\,\mathrm{mm}$
which is comparable to the gap width $L$, especially when the layer
resides on a well conducting metal plate. Finite extention of the
layer and the conducting bottom can partly be accounted for by a more
sophisticated complex potential, \[
w(\zeta)=\frac{\mu_{0}I_{0}}{2\pi}\log\frac{\sqrt{\zeta(\zeta+2H)}+\sqrt{L(L+2H)}}{\sqrt{\zeta(\zeta+2H)}-\sqrt{L(L+2H)}},\]
which is plotted in Fig. \ref{cap:edge-wire}(b). This yields $C_{0}=\mu_{0}I_{0}/\left\{ \pi\sqrt{L\left[1+L/(2H)\right]}\right\} ,$
resulting in $L\approx28\,\mathrm{mm}$ which is considerably closer
to the corresponding experimental value.

\begin{figure}
\begin{center}\includegraphics[%
  width=0.75\columnwidth]{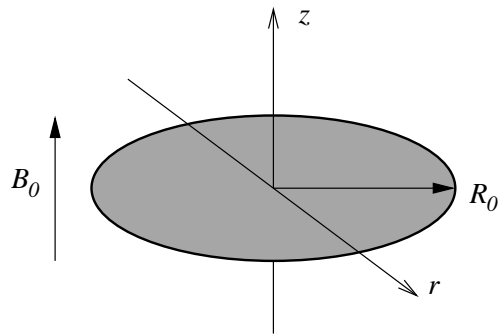}\end{center}

\caption{\label{cap:disk}Sketch of a thin perfectly conducting disk in axial
ac magnetic field.}
\end{figure}

\section{a thin liquid disk}

\subsection{Analytical solution}

\begin{figure}
\begin{center}\includegraphics[%
  bb=105bp 90bp 275bp 250bp,
  width=0.9\columnwidth]{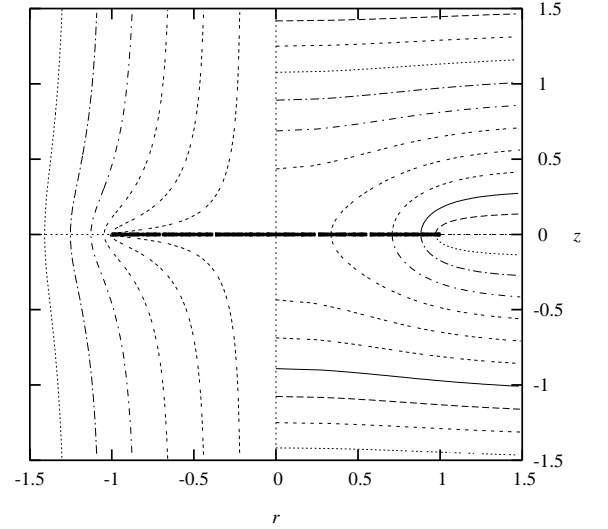}\end{center}

\caption{\label{cap:disk-Psi0}Isolines of the base magnetic potential $(r>0)$
and the corresponding magnetic flux lines $(r<0)$ around a circular
disk.}
\end{figure}

Now we will apply the approach developed in the previous section to
a thin circular liquid disk of radius $R_{0}$ and fixed thickness
$d_{0},$ which is subjected to a uniform axial ac magnetic field
with an induction amplitude $B_{0},$ as shown in Fig. \ref{cap:disk}.
The thickness $d_{0}$ is assumed to be small relative to the radius
of disk $d_{0}\ll R_{0},$ so that the disk may be regarded as a thin
sheet. The magnetic field is sought in terms of the scalar magnetic
potential $\Psi$ governed by Eq. (\ref{eq:Lapl-Psi}), whereas the
impermeability condition at the disk surface $S$ takes the form \begin{equation}
\left.\frac{\partial\Psi}{\partial n}\right|_{S}=0.\label{eq:bnd-disk}\end{equation}
 At large distances from the disk, the field is uniform and axial,
which implies\begin{equation}
\left.\Psi\right|_{\left|\vec{r}\right|\rightarrow\infty}\rightarrow(\vec{r}\cdot\vec{B}_{0})=zB_{0},\label{eq:infty-disk}\end{equation}
where $z$ is the axial distance from the disk. Solutions for both
a circular and a slightly perturbed disk can be obtained analytically
in the oblate spheroidal coordinates, which are related to the cylindrical
ones by\begin{eqnarray*}
r & = & R_{0}\sqrt{(1-\eta^{2})(1+\xi^{2})},\\
z & = & R_{0}\eta\xi,\end{eqnarray*}
where $0\le\eta\le1$ and $0\le\xi<\infty$ are the angular and radial
spheroidal coordinates, respectively, as defined in \cite{Li-etal-2002}
. Equation (\ref{eq:Lapl-Psi}) for the unperturbed potential $\Psi_{0}$
around a circular disk takes the form\begin{equation}
\frac{\partial}{\partial\eta}\left(\left(1-\eta^{2}\right)\frac{\partial\Psi_{0}}{\partial\eta}\right)+\frac{\partial}{\partial\xi}\left(\left(1+\xi^{2}\right)\frac{\partial\Psi_{0}}{\partial\xi}\right)=0.\label{eq:Lapl-Psi0-disk}\end{equation}
The impermeability condition (\ref{eq:bnd-disk}) is \begin{equation}
\left.\frac{\partial\Psi_{0}}{\partial\xi}\right|_{\xi=0}=0.\label{eq:bndc-disk}\end{equation}
 The second boundary condition (\ref{eq:infty-disk}) suggests a solution
of the form $\Psi_{0}(\eta,\xi)=\eta f_{0}(\xi).$ This results in
\cite{Bojarevics-etal}\begin{equation}
\Psi_{0}(\eta,\xi)=C_{0}\eta\left[1+\xi\arctan(\xi)\right],\label{eq:Psi0-dsk}\end{equation}
where $C_{0}=2R_{0}B_{0}/\pi,$ which is plotted in Fig. \ref{cap:disk-Psi0}
with the corresponding magnetic flux lines. Note that in the vicinity
of the edge this solution reduces to $\Psi_{0}=C_{0}\eta+O(\xi^{2}),$
which is equivalent to Eq. (\ref{eq:Psi0}).

\begin{figure}
\begin{center}\includegraphics[%
  bb=85bp 90bp 205bp 250bp,
  width=0.7\columnwidth,
  keepaspectratio]{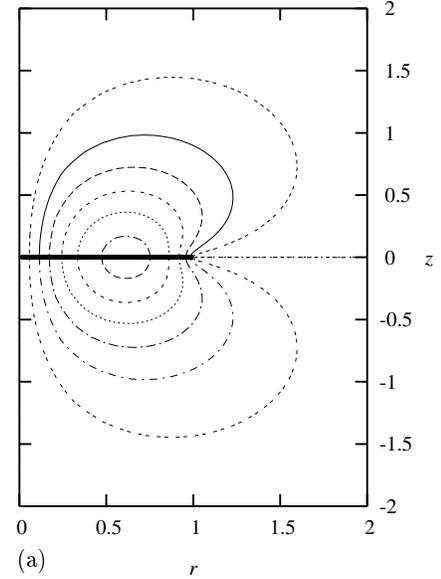}\put(-160,10){(a)}\end{center}

\begin{center}\includegraphics[%
  bb=85bp 90bp 205bp 250bp,
  width=0.7\columnwidth]{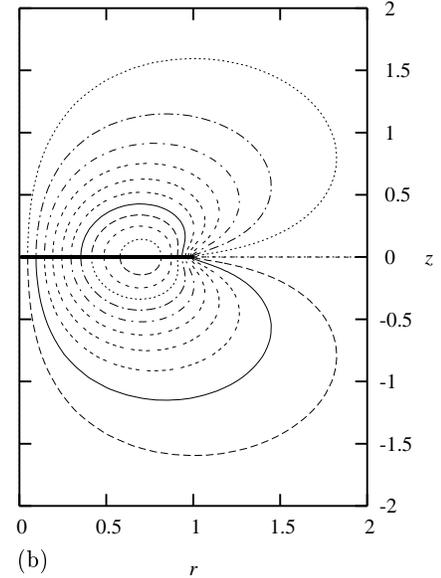}\put(-160,10){(b)}\end{center}

\caption{\label{cap:disk_Psi1} Isolines of perturbation amplitudes of the
magnetic potential relative to the perturbed edge for the azimuthal
modes $m=2$ (a) and $m=3$ (b) plotted with the step $0.03$ for
$C_{m}C_{0}=1.$ }
\end{figure}

Further, let us consider a perturbation of disk radius along the azimuthal
angle $\phi$ of the form\[
R(\phi,t)=R_{0}+R_{1}^{m}(\phi,t),\]
where $R_{1}^{m}(\phi,t)=\hat{R}_{1}^{m}(t)\cos(m\phi)$ is a small
perturbation with generally time-dependent amplitude $\hat{R}_{1}^{m}(t)$
for the wave number $m,$ which is integer in this case. Perturbation
of the disk disturbs the magnetic potential as \[
\Psi(\eta,\xi,\phi,t)=\Psi_{0}(\eta,\xi)+\hat{\Psi}_{1}^{m}(\eta,\xi,t)\cos(m\phi)+\cdots,\]
where $\hat{\Psi}_{1}^{m}$ is the perturbation amplitude, which is
associated with the wave number $m$, and satisfies the equation\begin{eqnarray}
\frac{\partial}{\partial\eta}\left(\left(1-\eta^{2}\right)\frac{\partial\hat{\Psi}_{1}^{m}}{\partial\eta}\right)+\frac{\partial}{\partial\xi}\left(\left(1+\xi^{2}\right)\frac{\partial\hat{\Psi}_{1}^{m}}{\partial\xi}\right)\nonumber \\
-\frac{m^{2}\left(\eta^{2}+\xi^{2}\right)}{\left(1-\eta^{2}\right)\left(1+\xi^{2}\right)}\hat{\Psi}_{1}^{m}=0\label{eq:Lapl-Psi1m-disk}\end{eqnarray}
 Perturbation of the magnetic potential is related to that of the
radius by the edge condition $\left.\Psi\right|_{r=R}=0$ resulting
in\begin{equation}
\left.\hat{\Psi}_{1}^{m}\right|_{\eta\rightarrow0}=-\hat{R}_{1}^{m}\left.\frac{\partial\Psi_{0}}{\partial r}\right|_{r\rightarrow R_{0}}=\frac{\hat{R}_{1}^{m}}{R_{0}}\left.\frac{C_{0}}{\eta}\right|_{\eta\rightarrow0}.\label{eq:disk-edge}\end{equation}
 This perturbation vanishes with the distance from the disk, i.e.,
$\left.\hat{\Psi}_{1}^{m}\right|_{\xi\rightarrow\infty}\rightarrow0.$
Although Eq. (\ref{eq:Lapl-Psi1m-disk}) admits variable separation,
such a solution is complicated by the edge singularity (\ref{eq:disk-edge}).
Nevertheless, a compact analytic solution can be found with the following
original approach. First, note that if $\Psi$ is a solution of the
Laplace equation and $\vec{\epsilon}$ is a constant vector, $(\vec{\epsilon}\cdot\vec{\nabla})\Phi$
is a solution too. Second, if $\Psi$ satisfies a uniform boundary
condition (\ref{eq:bnd-disk}) and $\vec{\epsilon}$ is directed along
the boundary, $(\vec{\epsilon}\cdot\vec{\nabla})\Phi$ satisfies that
boundary condition too. Third, the operator $(\vec{\epsilon}\cdot\vec{\nabla})$
changes the radial dependence of $\Psi$ from $\sim(r-R_{0})^{\alpha}$
to $\sim(r-R_{0})^{\alpha-1},$ while the azimuthal dependence is
changed from mode $m$ to $m+1.$ Algebra becomes particularly simple
when $\vec{\epsilon}$ is defined in the complex form as $\vec{\epsilon}=\vec{e}_{x}+i\vec{e}_{y}=e^{i\phi}(\vec{e}_{r}+i\vec{e}_{\phi}).$
Then each application of the complex operator $(\vec{\epsilon}\cdot\vec{\nabla})\Phi$
is accompanied by the multiplication with $e^{i\phi}.$ Thus, the
solution for $m=1$ follows simply from the axisymmetric basic state
as \begin{eqnarray*}
\hat{\Psi}_{1}^{1}(\eta,\xi) & = & -e^{-i\phi}C_{1}\left(\vec{\epsilon}\cdot\vec{\nabla}\right)\Psi_{0}\\
 & = & C_{1}C_{0}\left(\frac{1-\eta^{2}}{1+\xi^{2}}\right)^{1/2}\frac{\eta}{\eta^{2}+\xi^{2}}.\end{eqnarray*}
Similarly, higher azimuthal modes can be found as $\hat{\Psi}_{1}^{m}=e^{-im\phi}\left(\vec{\epsilon}\cdot\vec{\nabla}\right)^{m}\hat{\Psi}_{0}^{m},$
where $\hat{\Psi}_{0}^{m}$ is an axisymmetric solution satisfying
Eq. (\ref{eq:Lapl-Psi0-disk}). From the edge condition \[
\left(\vec{\epsilon}\cdot\vec{\nabla}\right)^{m}\Psi_{0}^{m}\sim\frac{\Psi_{0}^{m}}{\eta^{2m}}\sim\frac{1}{\eta}\]
 we obtain $\Psi_{0}^{m}\sim\eta^{2m-1}$ as $\eta\rightarrow0.$
Moreover, vanishing of perturbation far away from the disk $\left.\hat{\Psi}_{1}^{m}\right|_{\xi\rightarrow\infty}\rightarrow0$
implies that along the disk $\left.\hat{\Psi}_{0}^{m}\right|_{\xi=0}=c_{0}^{m}\eta^{2m-1},$
where $c_{0}^{m}$ is a constant. The corresponding axisymmetric solution
of Eq. (\ref{eq:Lapl-Psi0-disk}) can be represented as\[
\Psi_{0}^{m}(\eta,\xi)=c_{0}^{m}\sum_{k=1}^{m}c_{k}^{m}P_{2k-1}(\eta)Q_{2k-1}(i\xi),\]
where $P_{n}(x)$ and $Q_{n}(x)$ are the Legendre polynomials and
functions of the second kind, respectively \cite{Abramowitz}; the
expansion coefficients are $c_{k}^{m}=(4k-1)I_{k}^{m}/Q_{2k-1}(0),$
where\[
I_{k}^{m}=\int_{0}^{1}\eta^{2m-1}P_{2k-1}(\eta)d\eta=\frac{\sqrt{\pi}2^{1-2m}(2m-1)!}{(m-k)!\Gamma(m+k+1/2)}.\]
 Then the solution for perturbation amplitude can be written as \begin{equation}
\hat{\Psi}_{1}^{m}=D_{m-1}D_{m-2}\cdots D_{1}D_{0}\Psi_{0}^{m},\label{eq:Psi1m-dsk}\end{equation}
using the operator\[
D_{m}\equiv\frac{r}{R_{0}}\frac{1}{\eta^{2}+\xi^{2}}\left(-\eta\frac{\partial\,}{\partial\eta}+\xi\frac{\partial\,}{\partial\xi}\right)-\frac{mR_{0}}{r},\]
 which is a spectral analog of $\left(\vec{\epsilon}\cdot\vec{\nabla}\right)$
acting on the azimuthal mode $m.$ Calculation of Eq. (\ref{eq:Psi1m-dsk})
is algebraically complicated but can be done using \textsc{mathematica}
\cite{Mathematica}, which requires a considerable amount computer
memory and, thus, is possible only for $m\leq5.$ Nevertheless, this
suffices to deduce the general solution for arbitrary $m,$ \[
\hat{\Psi}_{1}^{m}(\eta,\xi)=C_{m}C_{0}\left(\frac{1-\eta^{2}}{1+\xi^{2}}\right)^{m/2}\frac{\eta}{\eta^{2}+\xi^{2}},\]
where the unknown constant $C_{m}=\hat{R}_{1}^{m}/R_{0}$ follows
from Eq. (\ref{eq:disk-edge}). It can easily be checked that the
above solution indeed satisfies both Eq. (\ref{eq:Lapl-Psi1m-disk})
and the edge condition (\ref{eq:disk-edge}) as well as the impermeability
condition (\ref{eq:bndc-disk}). As for the semi-infinite sheet, the
solution relative to the perturbed edge is obtained by the coordinate
transformation\begin{eqnarray*}
\Psi(\vec{r})=\Psi(\vec{r}'+\vec{e}_{r}R_{1}^{m}) & \approx & \Psi_{0}(\vec{r}')+\frac{\partial\Psi_{0}(\vec{r}')}{\partial r}R_{1}^{m}+\Psi_{1}^{m}(\vec{r}')\\
 & = & \Psi_{0}(\vec{r}')+\tilde{\Psi}_{1}^{m}(\vec{r}')\cos(m\phi),\end{eqnarray*}
 where $\tilde{\Psi}_{1}^{m}(\vec{r}')=\hat{\Psi}_{1}^{m}(\vec{r}')-\hat{\Psi}_{1}^{1}(\vec{r}').$
In the vicinity of the edge, this reduces to $\Psi(\eta,\xi)=\tilde{C}_{0}\eta+O(\xi^{2}),$
where $\tilde{C}_{0}=C_{0}\left[1-\frac{1}{2}(m-1)\hat{R}_{1}^{m}/R_{0}\right].$
Note that there is no perturbation of the magnetic field with respect
to the edge for $m=1,$ because this mode corresponds to the offset
of the disk as a whole. In this case, the field distribution moves
together with the disk causing perturbation with respect to the original
position of the disk, but not with respect to the disk itself. Perturbation
amplitudes $\tilde{\Psi}_{1}^{m}(\vec{r}')$ are plotted in Fig. \ref{cap:disk_Psi1}
for modes $m=2$ and $3$.

The time-averaged force on the edge follows from Eq. (\ref{eq:frc}),\[
F=F_{0}+F_{1}=F_{0}\left(1-(m-1)\frac{R_{1}^{m}}{R_{0}}\right),\]
where $F_{0}=\pi C_{0}^{2}/(8\mu_{0}R_{0})=B_{0}^{2}R_{0}/(2\pi\mu_{0}).$
Similarly to the semi-infinite sheet, the normal stress balance at
the edge of disk (\ref{eq:pressure}) results in\begin{equation}
-\rho\left.\frac{\partial\Phi}{\partial t}\right|_{r=R_{0}}=K_{1}\gamma+\frac{F_{1}}{l_{0}},\label{eq:balance-disk}\end{equation}
where $\gamma$ is the surface tension of the disk. The hydrodynamic
potential governed by Eq. (\ref{eq:Phi}) is found in cylindrical
coordinates as $\Phi(r,\phi,t)=\hat{\Phi}_{m}(t)r^{m}\cos(m\phi),$
while the kinematic constraint $v_{r}=\left.\partial\Phi/\partial r\right|_{r=R_{0}}=\partial R_{1}^{m}/\partial t$
yields $\hat{\Phi}_{m}(t)=1/(mR_{0}^{m-1})d\hat{R}_{1}^{m}/dt.$ The
curvature perturbation of the edge is \[
K_{1}=-\frac{R_{1}^{m}}{R_{0}^{2}}-\vec{\nabla}^{2}R_{1}^{m}=\frac{\hat{R}_{1}^{m}}{R_{0}^{2}}(m^{2}-1)\cos(m\phi).\]
 Searching the edge perturbation as $\hat{R}_{1}^{m}(t)=R_{1}e^{\lambda_{m}t},$
where $\lambda_{m}$ is in general a complex growth rate of the azimuthal
mode $m$, and substituting $\Phi$ and $K_{1}$ into Eq. (\ref{eq:balance-disk}),
we eventually obtain\begin{equation}
\lambda_{m}^{2}=\frac{m(m-1)}{\tau_{0}^{2}}\left(\frac{\Bm}{\pi}-m-1\right),\label{eq:disp-dsk}\end{equation}
 where $\tau_{0}=\sqrt{\rho R_{0}^{3}/\gamma}$ is the characteristic
time of capillary oscillations; $\Bm=B_{0}^{2}R_{0}^{2}/(2\mu_{0}l_{0}\gamma)$
is the dimensionless magnetic Bond number characterizing the ratio
of electromagnetic and surface tension forces. Without the magnetic
field $(\Bm=0),$ the growth rates for all modes are purely imaginary,
$\lambda_{m}=\pm i(m-1)\sqrt{m},$ corresponding to capillary oscillations
of an inviscid disk. Increasing the magnetic field results in a decrease
of the frequency of oscillations until the critical value of $\Bm$
is attained, at which an exponentially growing mode appears. According
to Eq. (\ref{eq:disp-dsk}), the critical Bond number for mode $m,$
which is defined by the condition $\lambda_{m}(\Bm_{c}^{m})=0$, is
$\Bm_{c}^{m}=(m+1)\pi,\, m=2,3,\ldots$. Note that for $m=0$ and
$1,$ we have $\lambda_{m}=0$ regardless of $\Bm$ because the first
mode is not permitted by the incompressibility constraint for the
layer of fixed thickness under consideration here. The mode $m=1,$
as already noted above, corresponds to the offset of the disk as a
whole, which has no effect relative to the disk itself as long as
the external magnetic field is uniform. Thus, the first unstable mode
is $m=2,$ for which the instability threshold is $\Bm_{c}=3\pi.$
Similarly to the straight edge case considered above, when $\Bm>\Bm_{c}$
the growth rate attains a maximum at the wave number $m_{\max}$ defined
by $\lambda_{m_{\max}}^{2}=\lambda_{m_{\max}-1}^{2}$ that yields
$m_{\max}=\left[\frac{2}{3}\Bm/\pi\right],$ where the square brackets
denote the integer part. 

\begin{figure}
\begin{center}\includegraphics[%
  width=0.6\columnwidth]{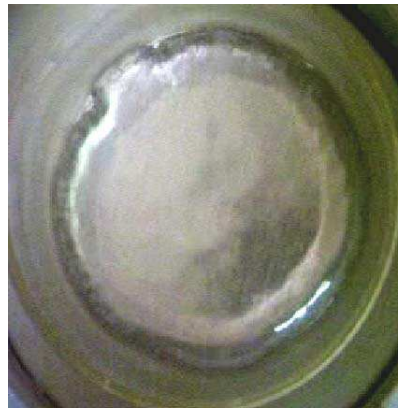}\put(-150,-10){(a)} \end{center}

\begin{center}\includegraphics[%
  width=0.6\columnwidth,
  keepaspectratio]{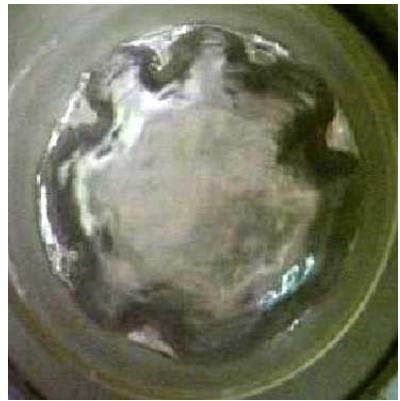}\put(-150,-10){(b)}\end{center}

\begin{center}\includegraphics[%
  width=0.6\columnwidth]{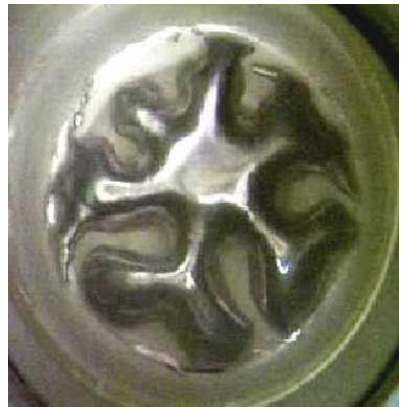}\put(-150,-10){(c)}\end{center}

\caption{\label{cap:exp}Top view of a flat Gallium drop in a transverse ac
magnetic field of $13\,\mathrm{kHz}$ frequency at various induction
amplitudes $B_{0}$: $12.9\,\mathrm{mT}$ (a), $23.5\,\mathrm{mT}$
(b), and $48.5\,\mathrm{mT}$ (c) .}
\end{figure}

\subsection{Comparison with experiment}

In the experiment, detailed description of which may be found in \cite{Perrier-etal-2003},
a flat circular gallium drop of thickness $d_{0}\approx6\,\mathrm{mm}$
and radius $R_{0}\approx30\,\mathrm{mm}$ was placed on a glass plate,
which was slightly concave to center the drop, and put into a 6-winding
solenoidal coil supplied by ac current of $f\approx13\,\mathrm{kHz}$
frequency. At low currents in the coil, the drop was observed to be
nearly circular, as seen in Fig. \ref{cap:exp}(a), and remained such
until the current exceeded some critical value, after which the drop
became noticeably distorted, as seen in Fig. \ref{cap:exp}(b). Further
current increase resulted in the development of more corrugated drop
shapes shown in Fig. \ref{cap:exp}(c). According to the experimental
observations, the circular shape became unstable about the magnetic
field induction amplitude in the range $12.9\,\mathrm{mT}<B_{0}<23.5\,\mathrm{mT}.$
Assuming the layer has approximately uniform curvature over the edge
thickness with the radius $r_{0}\approx d_{0}/2$ that corresponds
to an arclength $l_{0}\approx\pi d_{0}/2,$ we find $B_{0,c}=\sqrt{\pi\Bm_{c}\mu_{0}\gamma d_{0}}/R_{0}\approx13.4\,\mathrm{mT}$
for the critical Bond number $\Bm_{c}=3\pi,$ where $\gamma=0.718\,\mathrm{N}/\mathrm{m}$
is the surface tension of gallium \cite{Smithells}. This critical
field strength is slightly higher than that in Fig. \ref{cap:exp}(a)
but considerably lower than that in Fig. \ref{cap:exp}(b). For the
latter case we have $\Bm\approx30,$ which corresponds to the critical
wave number $m_{c}=\left[\Bm/\pi\right]-1=8$ defining the range of
linearly unstable modes $2\leq m\leq m_{c}.$ Note that the shape
seen in Fig. \ref{cap:exp}(b) has $m\approx8,$ which corresponds
to the critical mode for the given Bond number, although the fastest
growing mode in this case is $m_{\max}=6.$ Given the simplicity of
our theoretical model, these results may be thought to agree well
with the experiment. There may be several reasons that preclude a
better agreement with the experiment. First, the drop seen in Fig.
\ref{cap:exp}(b) has a significant perturbation amplitude implying
that its shape may be affected by nonlinear effects that are not accounted
for by this linear stability analysis. Second, our model of a thin
perfectly conducting sheet may be too simple for the given experiment
with the relative drop thickness $d_{0}/R_{0}\approx0.2$ and the
skin depth $\delta=1/\sqrt{\pi f\sigma\mu_{0}}\approx2.3\,\mathrm{mm},$
where $\sigma=3.7\times10^{6}\,\Omega^{-1}\mathrm{m}^{-1}$ is the
electrical conductivity of gallium \cite{Smithells}. 

Note that our theory is developed for a disk of fixed thickness which
excludes mode $m=0,$ while in the experiment the upper surface of
the layer is free and this mode is permitted. Nevertheless, the theory
is applicable also to this case because small-amplitude modes with
$m\geq1$ are not coupled with the mode $m=0.$ The only difference
is that the thickness of the layer may vary depending on the magnetic
field. But once the equilibrium thickness is known, our theory can
be used to predict whether the droplet will remain circular on the
further increase of the magnetic field.

\section{Summary and conclusions}

We have analyzed the linear stability of the edge of a thin liquid
metal layer subject to a transverse high-frequency ac magnetic field.
The metal layer was considered in the perfect conductor approximation
supposing the ac frequency to be high so that the magnetic field is
effectively expelled from the layer, while the thickness of the layer
was assumed to be small relative to its lateral extension so that
the layer could effectively be modeled as a thin perfectly conducting
liquid sheet. First, we considered a general model of a semi-infinite
sheet with a straight edge. This model, admitting an analytic solution,
allowed us to identify a pinch-type instability of the edge with the
following simple mechanism. The magnetic field bending around the
edge of a perfectly conducting layer creates a magnetic pressure on
the edge trying to compress the layer laterally. In the basic state
with a straight edge, the magnetic pressure, which is uniform along
the edge, is balanced by a constant hydrostatic pressure in the layer.
Perturbation of the edge in the form of a wave causes divergence of
magnetic flux lines at the wave crests and convergence in the troughs.
This redistribution is because the magnetic flux lines along the sheet
are perpendicular to the current lines. But since the latter are aligned
along the edge, the magnetic field has to be perpendicular to it.
Consequently, the magnetic pressure is reduced at the crests and increased
at the troughs, which drives the instability. Note that in this model
of a thin sheet, the induction varies with the distance $r$ from
the edge as $\sim1/\sqrt{r}$ and, thus, it formally becomes singular
at the edge. We circumvented this singularity by considering the sheet
to have a small but nevertheless finite thickness $d_{0}.$ Then integration
of the magnetic pressure, which scales as $\sim1/d_{0}$, over the
thickness $d_{0}$ resulted in a finite integral force on the edge
independent of its actual thickness. This allowed us to obtain an
analytical solution showing that the long-wave perturbations are unstable
when the wave number exceeds some critical value $k_{c}=F_{0}/(\gamma l_{0}),$
which is determined by the linear density of the electromagnetic force
$F_{0}$ acting on the edge, the surface tension $\gamma,$ and the
effective edge arclength $l_{0}$. The perturbations with wavelength
shorter than critical are stabilized by the surface tension, whereas
the growth rate of long-wave perturbations reduces as $\sim k$ for
$k\rightarrow0.$ Thus, there is the fastest growing perturbation
with the wave number $k_{\max}=2/3k_{c}.$ When the layer is arranged
vertically, long-wave perturbations are stabilized by the gravity,
and the critical perturbation is characterized by the capillary wave
number $k_{c}=\sqrt{g\rho/\gamma}.$ In this case, the critical linear
density of electromagnetic force is $F_{0,c}=2k_{c}l_{0}\gamma,$
which corresponds to the critical current amplitude $I_{0,c}=4\sqrt{\pi k_{c}l_{0}L\gamma/\mu_{0}}$
when the magnetic field is generated by a straight wire at the distance
$L$ directly above the edge. Next, we solved analytically the linear
stability problem for a thin circular disk placed in a transverse
uniform high-frequency ac magnetic field. It was found that the circular
shape of the disk becomes unstable with respect to exponentially growing
perturbation with the azimuthal wave number $m=2$ at the critical
magnetic Bond number $\Bm_{c}=3\pi.$ For $\Bm>\Bm_{c},$ the wave
number of the fastest growing perturbation is $m_{\max}=\left[2\Bm/(3\pi)\right].$
These theoretical results were found to be in reasonably good agreement
with available experimental data.

\begin{acknowledgments}
We thank Kirk Spragg for a critical reading of the manuscript. This
study was supported by the French-Latvian bilateral cooperation programme
in science and technology {}``Osmose'' under grant No. 06200PK.
J.P. gratefully acknowledges financial support from the Research Ministry
of France for senior researchers.
\end{acknowledgments}

\end{document}